\documentclass{aa}
\usepackage{graphicx}
\usepackage{txfonts}
\usepackage{natbib}

\newcommand{\tablenotea}[1]{\parbox{15.6cm}{ \indent
\footnotesize{\textsc{Note.--}~#1}}}
\newcommand{\tablenoteb}[1]{\parbox{8.6cm}{ \indent
\footnotesize{\textsc{Note.--}~#1}}}
\newcommand{\tablerefs}[1]{\parbox{8.6cm}{ \indent
\footnotesize{\textsc{References.--}~#1}}}

\newcommand{\cpl}{Chem.~Phys.~Lett.}

\newcommand{\jmsp}{J.~Mol.~Spectr.}
\newcommand{\jmst}{J.~Mol.~Struct.}
\newcommand{\nps}{Nature~Phys.~Sci.}
\newcommand{\aujp}{Aust.~J.~Phys.}
\newcommand{\znat}{Zeits.~Naturforsch.~A}
\newcommand{\fdis}{Faraday~Discuss.}
\begin{document}
\title{Detection of circumstellar CH$_2$CHCN, CH$_2$CN, CH$_3$CCH and
H$_2$CS\thanks{Based on observations carried out with the IRAM
30-meter telescope. IRAM is supported by INSU/CNRS (France), MPG
(Germany) and IGN (Spain).}}
\titlerunning{Circumstellar CH$_2$CHCN, CH$_2$CN, CH$_3$CCH and
H$_2$CS}
\authorrunning{Ag\'undez et al.}

\author{M. Ag\'undez\inst{1}, J. P. Fonfr\'ia\inst{1}, J. Cernicharo\inst{1}, J., R., Pardo\inst{1} \and M. Gu\'elin\inst{2}}

\offprints{M. Ag\'undez}

\institute{Departamento de Astrof\'isica Molecular e Infrarroja,
Instituto de Estructura de la Materia, CSIC, Serrano 121, E--28006
Madrid, Spain; \email{marce, pablo, cerni, pardo at
damir.iem.csic.es} \and Institut de Radioastronomie
Millim\'etrique, 300 rue de la Piscine, F--38406 St. Martin
d'H\'eres, France; \email{guelin@iram.fr}}

\date{Received ; accepted }


\abstract
{}
{We report on the detection of vinyl cyanide (CH$_2$CHCN),
cyanomethyl radical (CH$_2$CN), methylacetylene (CH$_3$CCH) and
thioformaldehyde (H$_2$CS) in the C-rich star IRC +10216. These
species, which are all known to exist in dark clouds, are detected
for the first time in the circumstellar envelope around an AGB
star.}
{The four molecules have been detected trough pure rotational
transitions in the course of a $\lambda$ 3 mm line survey carried
out with the IRAM 30-m telescope. The molecular column densities
are derived by constructing rotational temperature diagrams. A
detailed chemical model of the circumstellar envelope is used to
analyze the formation of these molecular species.}
{We have found column densities in the range 5 $\times$ 10$^{12}$
- 2 $\times$ 10$^{13}$ cm$^{-2}$, which translates to abundances
relative to H$_2$ of several 10$^{-9}$. The chemical model is
reasonably successful in explaining the derived abundances through
gas phase synthesis in the cold outer envelope. We also find that
some of these molecules, CH$_2$CHCN and CH$_2$CN, are most
probably excited trough infrared pumping to excited vibrational
states.}
{The detection of these species stresses the similarity between
the molecular content of cold dark clouds and C-rich circumstellar
envelopes. However, some differences in the chemistry are
indicated by the fact that in IRC +10216 partially saturated
carbon chains are present at a lower level than those which are
highly unsaturated, while in TMC-1 both types of species have
comparable abundances.}


\maketitle
%

\section{Introduction}

The chemical complexity of circumstellar envelopes (CSEs) around
carbon-rich AGB stars is illustrated by the well known object IRC
+10216, where more than 60 different molecular species have been
identified to date. Due to the carbon-rich nature of the gas, the
vast majority are oxygen-deprived hydrocarbons with a very
concrete structure consisting of a linear and highly unsaturated
backbone of carbon atoms. Among these species are cyanopolyynes
HC$_{2n+1}$N, polyacetylenic radicals C$_n$H, carbenes H$_2$C$_n$,
allenic radicals HC$_{2n}$N as well as sulphur and silicon-bearing
carbon chains C$_n$S, SiC$_n$ \citep{cer00}. The synthesis of
these molecules involves the photodissociation and photoionisation
by interstellar UV photons of parent species out-flowing from the
star and subsequent neutral-neutral and ion-molecule gas phase
reactions in the cold outer envelope (e.g.
\citealt{laf82,nej87,mil94,mil00}). Thus, reactive molecules are
distributed in circumstellar shells, as confirmed by
interferometric observations (e.g. \citealt{gue93}).

This type of chemistry resembles that occurring in cold dense
clouds, such as TMC-1, which are also rich in highly unsaturated
carbon chain molecules. The unsaturated character is typically the
result of low temperature non-equilibrium chemistry and reflects
the low reactivity of H$_2$ with most hydrocarbons at low
temperatures. Dark clouds contains however a sizable fraction of
partially saturated species, such as methylpolyynes
CH$_3$C$_{2n}$H (n=1,2 \citealt{irv81,wal84}), methylcyanopolyynes
CH$_3$C$_{2n+1}$N (n=0,1,2 \citealt{mat83ch3cn,bro84,sny06}),
cyanomethyl radical CH$_2$CN \citep{irv88}, vinyl cyanide
CH$_2$CHCN \citep{mat83ch2chcn}, cyanoallene CH$_2$CCHCN
\citep{lov06}; and highly saturated hydrocarbons such as propylene
CH$_2$CHCH$_3$ \citep{mar07}.

The question of why those partially saturated species are not
detected in IRC +10216 is a critical one for the understanding of
low temperature non-equilibrium chemistry. We have thus embarked
on a deep search for partially saturated organic molecules in IRC
+10216. In this paper we present the detection of CH$_2$CHCN,
CH$_2$CN and CH$_3$CCH, which are observed for the first time in a
CSE around an AGB star. The related species CH$_3$CN has been
already identified in this source by \citet{joh84}. Finally, we
also report on the detection of thioformaldehyde in IRC +10216.

\section{Spectroscopy and observations}

\begin{figure*}
\centering
\includegraphics[width=\textwidth]{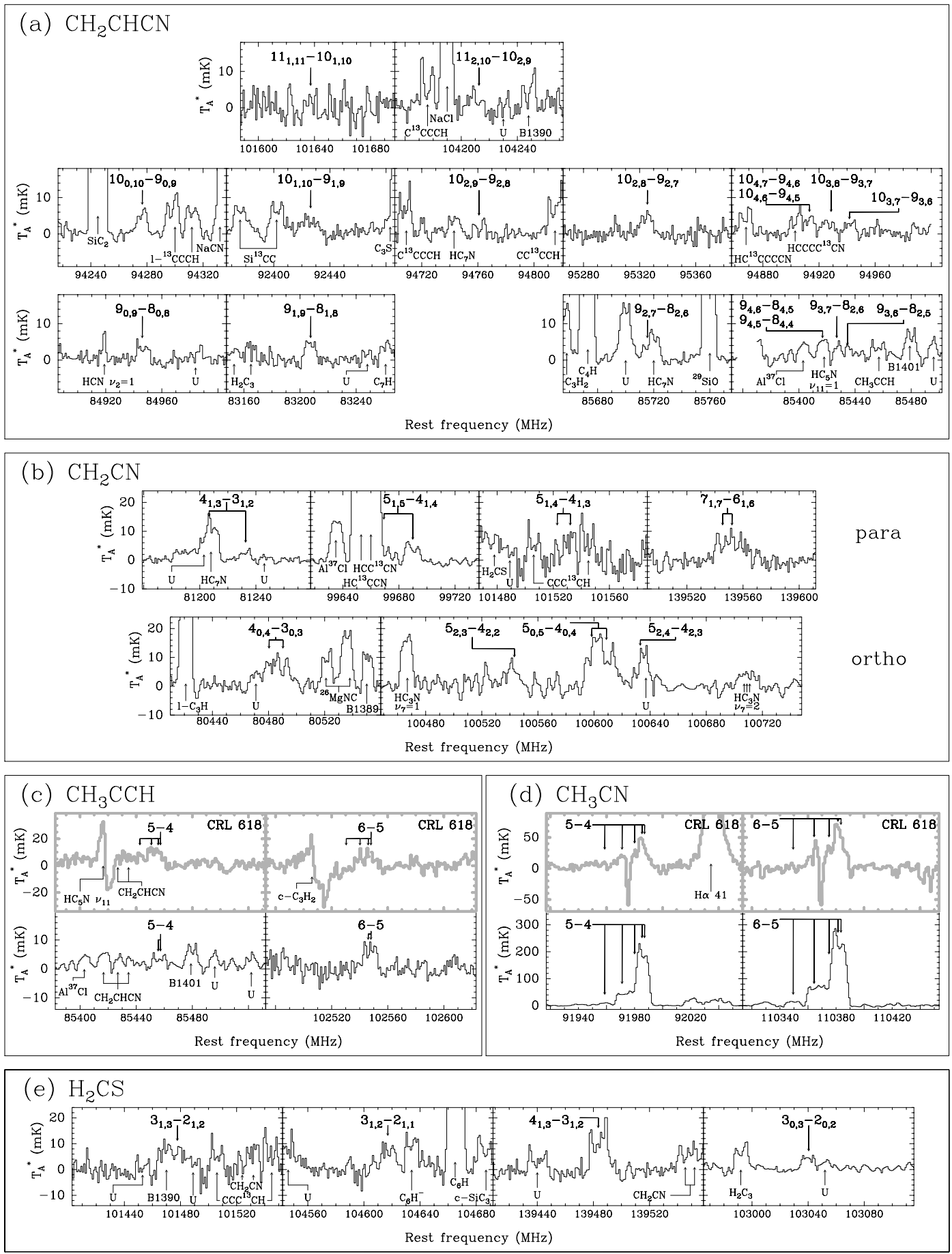}
\caption{Rotational lines of (a) CH$_2$CHCN, (b) CH$_2$CN, (c)
CH$_3$CCH, (d) CH$_3$CN and (e) H$_2$CS observed toward IRC +10216
with the IRAM 30-m telescope at a spectral resolution of 1 MHz.
The frequency scale is computed for an LSR source velocity of
-26.5 km s$^{-1}$. For comparison purposes, the thick grey panels
show spectra corresponding to CRL 618 (from \citealt{par07}).
``U'' means Unidentified line.} \label{fig-lines}
\end{figure*}

\begin{table*}
\caption{Line parameters of CH$_2$CHCN, CH$_2$CN, CH$_3$CCH and
H$_2$CS in IRC +10216} \label{table-lineparameters} \centering
\begin{tabular}{ccrrcrrrc}
\hline \hline
\multicolumn{2}{c}{} & \multicolumn{1}{c}{Observed} &\multicolumn{1}{c}{Calculated} & \multicolumn{1}{c}{} & \multicolumn{1}{c}{Line} & \multicolumn{3}{c}{} \\
\multicolumn{1}{c}{Transition} & \multicolumn{1}{c}{Symmetry} & \multicolumn{1}{c}{Frequency} &\multicolumn{1}{c}{Frequency} & \multicolumn{1}{c}{v$_{exp}$$^a$} & \multicolumn{1}{c}{Strength} & \multicolumn{1}{c}{E$_{up}$} & \multicolumn{1}{c}{$\int$T$_{A}^*$dv} & \multicolumn{1}{c}{$\eta_b$$^b$} \\
\multicolumn{2}{c}{} & \multicolumn{1}{c}{(MHz)} & \multicolumn{1}{c}{(MHz)} &\multicolumn{1}{c}{(km s$^{-1}$)} & \multicolumn{1}{c}{} & \multicolumn{1}{c}{(K)} & \multicolumn{1}{c}{(K km s$^{-1}$)} & \multicolumn{1}{c}{} \\
\hline
\multicolumn{9}{c}{CH$_2$CHCN} \\
\hline
9$_{0,9}$-8$_{0,8}$              &    & 84946.7(3)   & 84946.000    & 15.5(10) & 9.00     & 20.4     & 0.11(2)     & 0.82 \\
9$_{1,9}$-8$_{1,8}$              &    & 83207.5(8)   & 83207.505    & 14.8(10) & 8.89     & 22.1     & 0.14(2)     & 0.82 \\
9$_{2,7}$-8$_{2,6}$              &    & 85715.4(5)   & 85715.424    & 14.5$^d$ & 8.56     & 29.2     & 0.09(3)$^c$ & 0.82 \\
9$_{3,7}$-8$_{3,6}$              &    & 85427.9(10)  & 85426.920    & 14.5$^d$ & 8.00     & 40.0     & 0.09(3)$^c$ & 0.82 \\
9$_{3,6}$-8$_{3,5}$              &    & 85435.6(10)  & 85434.526    & 14.5$^d$ & 8.00     & 40.0     & 0.07(2)$^c$ & 0.82 \\
10$_{0,10}$-9$_{0,9}$            &    & 94276.3(10)  & 94276.634    & 14.5$^d$ & 9.99     & 25.0     & 0.15(3)     & 0.81 \\
10$_{1,10}$-9$_{1,9}$            &    & 92426.1(15)  & 92426.248    & 14.5$^d$ & 9.90     & 26.6     & 0.11(3)     & 0.81 \\
10$_{2,9}$-9$_{2,8}$             &    & 94760.9(7)   & 94760.781    & 14.5$^d$ & 9.60     & 33.7     & 0.07(2)     & 0.81 \\
10$_{2,8}$-9$_{2,7}$             &    & 95325.1(8)   & 95325.474    & 14.5$^d$ & 9.60     & 33.8     & 0.14(3)     & 0.81 \\
10$_{3,8}$-9$_{3,7}$             &    & 94926.2(15)  & 94928.606    & 14.5$^d$ & 9.10     & 44.5     & 0.09(3)     & 0.81 \\
10$_{3,7}$-9$_{3,6}$             &    & 94942.0(10)  & 94941.630    & 14.5$^d$ & 9.10     & 44.5     & 0.08(2)     & 0.81 \\
10$_{4,7}$-9$_{4,6}$             &    & 94913.4(10)  & 94913.115    & 14.5$^d$ & 8.40     & 59.7     & 0.11(3)$^c$ & 0.81 \\
10$_{4,6}$-9$_{4,5}$             &    &              & 94913.226    &          & 8.40     & 59.7     & $^c$        & 0.81 \\
11$_{1,11}$-10$_{1,10}$          &    & 101637.1(8)  & 101637.231   & 14.5$^d$ & 10.9     & 31.5     & 0.07(3)     & 0.80 \\
11$_{2,10}$-10$_{2,9}$           &    & 104211.5(10) & 104212.646   & 14.5$^d$ & 10.6     & 38.7     & 0.08(3)     & 0.79 \\
\hline
\multicolumn{9}{c}{CH$_2$CN} \\
\hline
4$_{0,4}$-3$_{0,3}$ J=9/2-7/2    & O  & 80480.5(10)  & 80480.384    & 14.5$^d$ & 4.44     & 9.7      & 0.23(6)$^c$ & 0.82 \\
4$_{0,4}$-3$_{0,3}$ J=7/2-5/2    & O  & 80489.6(10)  & 80490.261    & 14.5$^d$ & 3.43     & 9.7      & 0.25(6)$^c$ & 0.82 \\
5$_{0,5}$-4$_{0,4}$ J=11/2-9/2   & O  & 100600.2(15) & 100598.383   & 14.5$^d$ & 5.46     & 14.5     & 0.40(7)$^c$ & 0.80 \\
5$_{0,5}$-4$_{0,4}$ J=9/2-7/2    & O  & 100609.6(15) & 100608.832   & 14.5$^d$ & 4.44     & 14.5     & 0.33(6)$^c$ & 0.80 \\
5$_{2,4}$-4$_{2,3}$ J=11/2-9/2   & O  & 100633.2(15) & 100632.957   & 14.5$^d$ & 4.58     & 67.2     & 0.10(3)$^c$ & 0.80 \\
5$_{2,3}$-4$_{2,2}$ J=9/2-7/2    & O  & 100542.5(10) & 100543.214   & 14.5$^d$ & 3.73     & 67.2     & 0.12(3)     & 0.80 \\
4$_{1,3}$-3$_{1,2}$ J=7/2-5/2    & P  & 81207.6(10)  & 81206.601    & 14.5$^d$ & 3.22     & 8.9      & 0.09(3)$^c$ & 0.82 \\
4$_{1,3}$-3$_{1,2}$ J=9/2-7/2    & P  & 81232.4(5)   & 81232.654    & 14.8(5)  & 4.17     & 8.9      & 0.09(2)     & 0.82 \\
5$_{1,5}$-4$_{1,4}$ J=11/2-9/2   & P  & 99689.5(8)   & 99689.833    & 14.5$^d$ & 5.23     & 13.4     & 0.14(4)     & 0.80 \\
5$_{1,4}$-4$_{1,3}$ J=11/2-9/2   & P  & 101531.6(10) & 101532.055   & 14.5$^d$ & 5.23     & 13.7     & 0.22(6)     & 0.80 \\
7$_{1,7}$-6$_{1,6}$ J=13/2-11/2  & P  & 139547.1(15) & 139545.477   & 14.5$^d$ & 6.33     & 25.9     & 0.14(4)     & 0.75 \\
7$_{1,7}$-6$_{1,6}$ J=15/2-13/2  & P  & 139553.4(10) & 139552.138   & 14.5$^d$ & 7.30     & 25.9     & 0.12(4)     & 0.75 \\
\hline
\multicolumn{9}{c}{CH$_3$CCH} \\
\hline
J$_K$=5$_0$-4$_0$                &    &  85456.4(10) &  85457.299   & 14.5$^d$ & 5.00     & 11.5     & 0.12(4)$^c$ & 0.82 \\
J$_K$=5$_1$-4$_1$                &    &              &  85455.665   &          & 4.80     & 20.3     & $^c$        & 0.82 \\
J$_K$=6$_0$-5$_0$                &    & 102547.5(6)  & 102547.983   & 14.8(6)  & 6.00     & 17.2     & 0.22(5)$^c$ & 0.80 \\
J$_K$=6$_1$-5$_1$                &    &              & 102546.023   &          & 5.83     & 24.4     & $^c$        & 0.80 \\
\hline
\multicolumn{9}{c}{H$_2$CS} \\
\hline
3$_{1,3}$-2$_{1,2}$              & O  & 101478.1(10) & 101477.750   & 14.5$^d$ & 2.67     & 8.1      & 0.18(4)$^c$ & 0.80 \\
3$_{1,2}$-2$_{1,1}$              & O  & 104617.5(10) & 104616.969   & 14.5$^d$ & 2.67     & 8.4      & 0.23(3)     & 0.79 \\
4$_{1,3}$-3$_{1,2}$              & O  & 139483.8(5)  & 139483.422   & 14.1(6)  & 3.75     & 15.1     & 0.36(5)     & 0.75 \\
3$_{0,3}$-2$_{0,2}$              & P  & 103040.3(5)  & 103040.396   & 14.6(7)  & 3.00     & 9.9      & 0.13(2)     & 0.80 \\
\hline
\end{tabular}
\tablenotea{The numbers in parentheses are errors in units of the
last digits; the 1$\sigma$ error in $\int$T$_{A}^*$dv is derived
from least-squares fits and does not include the calibration
uncertainty of 10 \%. $^a$ v$_{exp}$ stands for expansion velocity
and is computed as half the full linewidth at zero level. $^b$
$\eta_b$ is B$_{\rm eff}$/F$_{\rm eff}$, i.e. the ratio of T$_A^*$
to T$_{mb}$. A superscript ``c'' denotes a partly blended line
while ``d'' indicates that the linewidth parameter v$_{exp}$ has
been fixed to 14.5 km s$^{-1}$. The calculated frequencies of
CH$_2$CN correspond to the strongest hyperfine component, as
measured by \citet{oze04}. CH$_2$CN and H$_2$CS lines have ortho
(O) or para (P) symmetry; E$_{up}$ is the energy of the upper
level of the transition above the O or P ground state.}
\end{table*}

The four molecules on focus in this paper (CH$_2$CHCN, CH$_2$CN,
CH$_3$CCH and H$_2$CS) were identified in space for the first time
towards Sagittarius B2 \citep{gar75,irv88,sny73,sin73} by
observation of rotational transitions. The microwave spectrum of
all these species has been extensively studied in the laboratory
so that their spectroscopic properties are accurately known; see a
compilation in the JPL and Cologne databases for molecular
spectroscopy \citep{pic98,mul05}.

CH$_2$CHCN is a planar asymmetric rotor with $a$- and $b$-type
allowed transitions ($\mu_a$ = 3.815 D and $\mu_b$ = 0.894 D
;\citealt{sto85}). All the transitions observed in IRC +10216
belong to the stronger $a$-type.

The radical CH$_2$CN has two interchangeable hydrogen nuclei with
non zero spin which result in two distinct groups of rotational
levels: ortho (parallel spins) and para (antiparallel spins) with
relative statistical weights 3:1, between which both radiative and
collisional transitions are highly forbidden. The electronic
ground state is $^2B_1$, thus, the quantum number $K_a$ is even
for ortho levels and odd for para levels. The para ground state
(1$_{1,1}$) lyes 14.15 K above the ortho ground state (0$_{0,0}$).
This assignment is reversed compared to more common species with a
$^1A_1$ electronic ground state (H$_2$CO, H$_2$CS, ...), where the
quantum number $K_a$ is odd for ortho levels and even for para
levels, with the rotational ground state having para symmetry. Its
dipole moment is relatively large: $\mu_a$ = 3.5 D \citep{oze04}.
The unpaired electron causes spin-rotation coupling which splits
each transition
$N^{'}_{K^{'}_{a},K^{'}_{c}}$-$N^{''}_{K^{''}_{a},K^{''}_{c}}$
into two components. Further hyperfine coupling of rotation with
the non-zero spin of $^1$H and $^{14}$N nuclei produces a myriad
of components whose frequencies have been precisely measured in
the laboratory \citep{oze04}.

Propyne (CH$_3$CCH) is a prolate symmetric top molecule whose
rotational levels are divided into two different species, A-type
and E-type, not connected radiatively. Its dipole moment is
relatively small: $\mu_a$ = 0.784 D \citep{bur80}. The rotational
spectrum was first measured by \citet{tra50} and is now known with
a high accuracy for its ground and some excited vibrational states
\citep{mul02}.

H$_2$CS is an asymmetric rotor which, analogously to CH$_2$CN, has
two interchangeable hydrogen nuclei so that its rotational levels
are grouped into ortho ($K_a$ odd) and para ($K_a$ even), with
statistical weights 3:1. The ortho ground state (1$_{1,1}$) lyes
14.9 K above the para ground state (0$_{0,0}$). Its dipole moment
was measured by \citet{fab77} to be $\mu_a$ = 1.647 D.

The observations shown in Fig.~\ref{fig-lines} were made with the
IRAM 30-m telescope during several sessions from 1990 to 2006,
most of them after 2002 in the context of a $\lambda$ 3 mm line
survey of IRC +10216 (Cernicharo et al. in preparation). SIS
receivers operating at 3 and 2 mm were tuned in single sideband
mode, with typical image rejections larger than 20 dB at 3 mm and
around 15 dB at 2 mm. We express intensities in terms of T$_A^*$,
the antenna temperature corrected for atmospheric absorption and
for antenna ohmic and spillover losses. The uncertainty in T$_A^*$
due to calibration is estimated to be around 10 \%. Data were
taken in the standard wobbler switching mode with a beam throw of
4'. Pointing and focusing were checked by observing nearby planets
and the quasar OJ 287. The back end used was a filterbank with a
bandwidth of 512 MHz and a spectral resolution of 1.0 MHz. The
system temperature was 100-150 K at 3 mm and 200 K at 2 mm. On
source integration times ranged from 2 to 20 hours, resulting in
rms noise levels of 1-3 mK in T$_A^*$.

\section{Results}

Fig.~\ref{fig-lines} shows the observed spectra of IRC +10216
corresponding to the CH$_2$CHCN, CH$_2$CN, CH$_3$CCH and H$_2$CS
lines. The observational line parameters, obtained by using the
SHELL fitting routine of the CLASS
package\footnote{http://www.iram.fr/IRAMFR/GILDAS}, are given in
Table~\ref{table-lineparameters}. Linewidths have been fixed in
many cases due to limited signal-to-noise (S/N) ratio or due to
blending with other lines. The measured widths are consistent with
an expansion velocity v$_{exp}$ around 14.5 km s$^{-1}$, in
agreement with most of the molecular lines arising from the
expanding envelope of IRC +10216 \citep{cer00}. There are no
missing lines of CH$_2$CN, CH$_3$CCH or H$_2$CS in the 3 mm
atmospheric window. However, some lines of CH$_2$CHCN having
similar strengths and upper level energies to those reported in
Table~\ref{table-lineparameters} are not listed due to a complete
blending with a stronger line or due to a low sensitivity of the
spectra. For example the 11$_{0,11}$-10$_{0,10}$ transition at
103575 MHz is blended with a strong component of the $J$=21/2-19/2
$^2\Pi_{1/2}$ $\nu_7$=1 doublet of C$_4$H.

The line profiles can give us information on the spatial
distribution of the molecules. In a spherically expanding envelope
and for lines with a low optical depth, a double-peaked shape
indicates that the emitting species has a distribution with an
angular extent comparable or larger than the telescope beam (HPBW
= 21''-31'' for IRAM 30-m at $\lambda$ 3 mm), while a flat-topped
profile indicates that the emitting region is not spatially
resolved by the telescope beam. The lines of CH$_3$CN in IRC
+10216 have a double-peaked character (see Fig.~\ref{fig-lines}),
which indicates that this molecule has an extended distribution.
In the case of the four species with which we are concerned, the
low S/N ratio of most of the observed lines makes difficult to
distinguish whether line profiles are predominantly double-peaked
or flat-topped for a given species. Therefore, any conclusion
about the spatial distribution of the molecules derived from the
line profiles must be taken with caution. We will, nevertheless,
consider that CH$_2$CHCN, CH$_2$CN, CH$_3$CCH and H$_2$CS have an
extended distribution, based on chemical arguments (see \S 3.2 and
Fig.~\ref{fig-chem}), which essentially fills the telescope beam.
Under such hypothesis, we have constructed rotational temperature
diagrams to derive beam averaged column densities.

\begin{figure}
\includegraphics[angle=0,scale=.53]{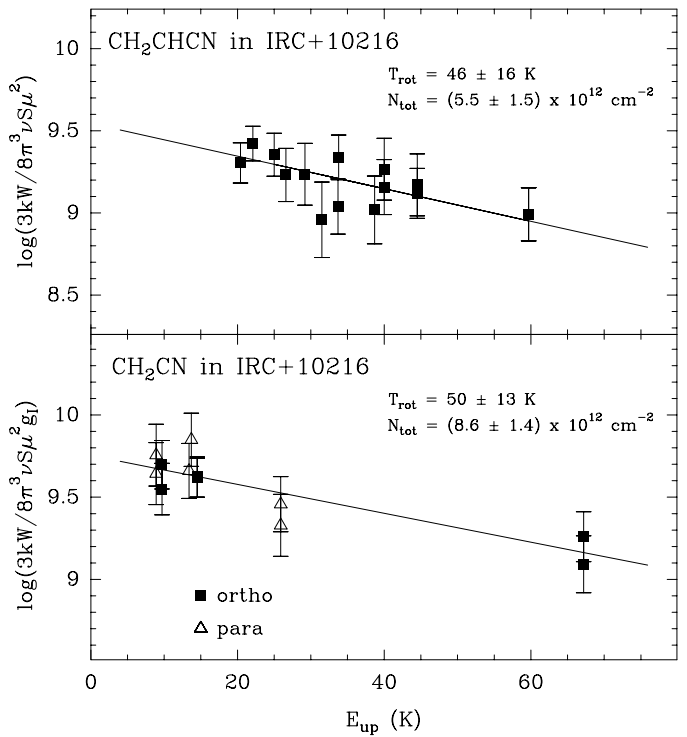}
\caption{Rotational temperature diagrams of CH$_2$CHCN and
CH$_2$CN in IRC +10216. T$_{rot}$ and N$_{tot}$ have been
determined from a weighted least-squares fit to the observational
points, and their uncertainties reflect the 1$\sigma$ errors in
the fits to the line profiles plus the 10 \% estimated uncertainty
in the calibration.} \label{fig-rtd}
\end{figure}

For CH$_2$CHCN (see Fig.~\ref{fig-rtd}) we derive a total column
density N$_{tot}$ = (5.5 $\pm$ 1.5) $\times$ 10$^{12}$ cm$^{-2}$
and a rotational temperature of 46 $\pm$ 16 K, which is within the
range of rotational temperatures derived for other
shell-distributed molecules in IRC +10216: 20-50 K \citep{cer00}.

In the case of CH$_2$CN, we have used all the observed lines
--ortho and para-- in a single rotational diagram (see
Fig.~\ref{fig-rtd}). Given that the lines are detected with a low
S/N ratio, we do not aim at determining the O/P ratio although the
available data is consistent with the statistical value 3:1. We
derive a rotational temperature of 50 $\pm$ 13 K and a total
column density of N$_{tot}$ = (8.6 $\pm$ 1.4) $\times$ 10$^{12}$
cm$^{-2}$, which is comparable to that of the related species
CH$_3$CN (see Table~\ref{table-columndensities}).

The two CH$_3$CCH lines detected are very likely a sum of the $K$
= 0, 1 components, although the low S/N ratio do not allow to
clearly distinguish them. Assuming a rotational temperature of 30
K we tentatively derive a total column density of 1.8 $\times$
10$^{13}$ cm$^{-2}$. This value is comparable to the column
density of CH$_3$CN in IRC +10216, which is detected through lines
about 25 times stronger than those of CH$_3$CCH due to its much
larger dipole moment ($\mu_{CH_3CN}$ = 3.925 D versus
$\mu_{CH_3CCH}$ = 0.780 D).

Both CH$_3$CN and CH$_3$CCH have been also detected in the C-rich
preplanetary nebula CRL 618 (\citealt{cer01,par06,par07}; see also
Fig.\ref{fig-lines}), which is in a evolutionary stage immediately
following that of IRC +10216. However, the CH$_3$CCH/CH$_3$CN
ratio is noticeably different in these two objects. In IRC +10216
both molecules are present with a similar abundance
(CH$_3$CCH/CH$_3$CN = 0.6). In contrast, in CRL 618
methylacetylene is much more abundant than methyl cyanide
\citep{par07}, not only in the region corresponding to the first
post-AGB ejections (CH$_3$CCH/CH$_3$CN = 6) but also in the warm
and dense inner regions where UV photons efficiently drive a rich
C-based photochemistry (CH$_3$CCH/CH$_3$CN = 15). Such a stage has
not yet been reached by IRC +10216.

The four lines of H$_2$CS observed in IRC +10216 have similar
E$_{up}$ values, thus it is rather difficult to constrain the
rotational temperature. Assuming an statistical O/P ratio of 3:1
and a rotational temperature of 30 K we derive a total column
density N$_{tot}$ = 1 $\times$ 10$^{13}$ cm$^{-2}$ for
thioformaldehyde, which is almost twice larger than that of
formaldehyde (\citealt{for04}; see
Table~\ref{table-columndensities}).

\section{Chemical modelling}

\begin{figure}
\includegraphics[angle=-90,scale=.41]{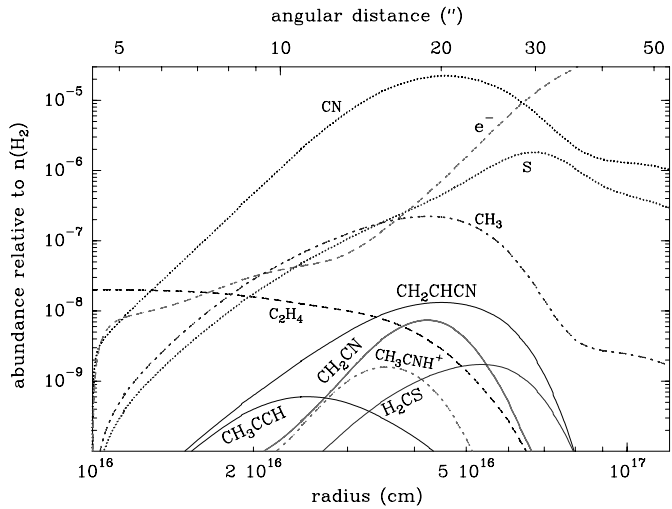}
\caption{Abundances of CH$_2$CHCN, CH$_2$CN, CH$_3$CCH and H$_2$CS
(solid lines) and related species (dotted and dashed lines) given
by the chemical model, as a function of radius (bottom axis) and
angular distance (top axis) for an assumed stellar distance of 150
pc.} \label{fig-chem}
\end{figure}

In order to explain how the detected species are formed we have
constructed a detailed chemical model of the outer envelope. The
chemical network consists of 385 gas phase species linked by 6547
reactions, whose rate constants have been taken from
\citet{cer04}, \citet{agu06} and from the UMIST Database for
Astrochemistry \citep{woo07}. The temperature and density radial
profiles as well as other physical assumptions are taken from
\citet{agu06}. The resulting abundance radial profiles for
CH$_2$CHCN, CH$_2$CN, CH$_3$CCH, H$_2$CS and related species are
plotted in Fig.~\ref{fig-chem}. The model predicts that these four
molecules would display an extended shell-type distribution with
an angular radius of about 20''. The predicted and observed column
densities are summarized in Table~\ref{table-columndensities}.

Vinyl cyanide results from the reaction between CN and ethylene,
the latter being found in the inner envelope with an abundance
relative to H$_2$ of 2 $\times$ 10$^{-8}$ \citep{gol87}. This
reaction has been studied in the laboratory and found to be very
rapid at low temperature \citep{sim93} yielding vinyl cyanide
\citep{cho04}:
\begin{equation}
CN + C_2H_4 \longrightarrow CH_2CHCN + H
\end{equation}
Our model predicts a total column density N$_{tot}$ = 1.4 $\times$
10$^{13}$ cm$^{-2}$, which reasonably agrees with the value
derived from the observations\footnote{The radial column densities
given by the model are multiplied by a factor 2 to get the total
column densities across the envelope N$_{tot}$.}. The analogue of
reaction (1) replacing C$_2$H$_4$ by C$_2$H$_2$ is the main
formation pathway of HC$_3$N in IRC +10216. The much larger column
density of HC$_3$N compared to that of CH$_2$CHCN (more than two
orders of magnitude; see Table~\ref{table-columndensities})
reflects the large excess of acetylene over ethylene in the inner
envelope of IRC +10216. In dark clouds, reaction (1) is also
probably the main formation route to vinyl cyanide \citep{her90}.

\begin{table}
\caption{Column densities of selected molecules in IRC +10216}
\label{table-columndensities} \centering
\begin{tabular}{llc@{ }c@{ }ccc}
\hline \hline
\multicolumn{1}{l}{Molecule} & \multicolumn{1}{c}{T$_{rot}$ (K)} & \multicolumn{5}{c}{N$_{tot}$ (cm$^{-2}$)}    \\
\hline
                    &                     & \multicolumn{3}{c}{observed} & \multicolumn{2}{c}{calculated} \\
\cline{3-7}
                    &                     & & & & \multicolumn{1}{c}{This Work} & \multicolumn{1}{c}{[1]} \\
\cline{6-7}
HC$_3$N    & 15     & & 2.2(15) & $^\textnormal{\scriptsize{[2]}}$ & 2.8(15) & 3.6(15) \\
\multicolumn{7}{c}{} \\
\textbf{CH$_2$CHCN} & 46     & & 5.5(12) &      & 1.4(13) & 2.2(11) \\
\multicolumn{7}{c}{} \\
\textbf{CH$_2$CN}   & 50     & & 8.6(12) &      & 4.6(12) & 1.4(13) \\
CH$_3$CN            & 40     & & 3.0(13) & $^\textnormal{\scriptsize{[2]}}$       & 6.3(12) & 6.8(12) \\
CH$_3$C$_3$N        & 30 $^a$ & $<$ & 1.3(12) & $^\textnormal{\scriptsize{[2]}}$    & 1.2(11) & 1.4(12) \\
\textbf{CH$_3$CCH}  & 30 $^a$ & & 1.8(13) &     & 1.1(12) & 8.0(12) \\
CH$_3$C$_4$H        & 30 $^a$ & $<$ & 9.7(12) & $^\textnormal{\scriptsize{[2]}}$    & 8.2(12) & 9.0(12) \\
\multicolumn{7}{c}{} \\
H$_2$CO             & 28     & & 5.4(12) & $^\textnormal{\scriptsize{[3]}}$ & 2.8(12) & --      \\
\textbf{H$_2$CS}    & 30 $^a$ & & 1.0(13) &      & 1.3(12) & 4.4(11) \\
\hline
\end{tabular}
\tablenoteb{a(b) refers to a$\times$10$^b$. A superscript ``a''
indicates an assumed rotational temperature.} \tablerefs{[1]:
\citet{mil00}; [2]: from unpublished IRAM 30-m data; [3]: from the
observations by \citet{for04}.}
\end{table}

The dissociative recombination (DR) of the ion CH$_3$CNH$^+$ is
the major pathway to form both CH$_2$CN and CH$_3$CN:
\begin{equation}
\begin{array}{rl}
CH^+ \displaystyle{\stackrel{H_2}{\longrightarrow}} CH_2^+ \displaystyle{\stackrel{H_2}{\longrightarrow}} CH_3^+ \displaystyle{\stackrel{HCN}{\longrightarrow}} CH_3CNH^+ & \displaystyle{\stackrel{e^-}{\longrightarrow}} CH_2CN \\
 & \displaystyle{\stackrel{e^-}{\longrightarrow}} CH_3CN
\end{array}
\end{equation}
The branching ratios of the channels giving CH$_2$CN and CH$_3$CN
are not known and are assumed to be equal. In our model, the major
destruction process for both CH$_2$CN and CH$_3$CN is the
photodissociation by interstellar UV photons. The
photodissociation rate of CH$_3$CN has been calculated by
\citet{rob91} but that of CH$_2$CN is unknown and we assume it to
be the same as that of CH$_3$CN. Since the chemistry of these two
molecules is relatively well constrained in IRC +10216, they are
formed by the same reaction and their destruction is dominated by
just one process, we may use the observed CH$_2$CN/CH$_3$CN ratio
to estimate the branching ratios of the DR of the CH$_3$CNH$^+$
ion. We find an 80 \% for the channel (CH$_3$CN + H) and 20 \% for
(CH$_2$CN + H$_2$ or 2H). This estimate will be strongly affected
if the photodissociation rate of CH$_2$CN is very different from
that of CH$_3$CN but not if, as suggested by \citet{her90} and
\citet{tur90}, CH$_2$CN reacts with atomic oxygen, the abundance
of which is low in IRC +10216 at the radius where CH$_2$CN is
present.

The related species CH$_3$C$_3$N is formed by the same sequence of
reactions that produces CH$_3$CN, just replacing HCN by HC$_3$N.
Its predicted column density is lower than that of CH$_3$CN by
more than one order of magnitude. However, since our model
predicts less CH$_3$CN than observed, the CH$_3^+$ ion is probably
underproduced, the column density of CH$_3$C$_3$N is most likely
underestimated. If CH$_3$C$_3$N was present with a column density
of several 10$^{11}$ cm$^{-2}$, the expected brightness
temperatures would be a few mK and therefore it could be
detectable. In fact, our IRAM 30-m line survey at $\lambda$ 3 mm
of IRC +10216 (Cernicharo et al. in preparation) shows two
unidentified features with $T_A^*$ $\sim$2-3 mK at 86756 MHz and
103280 MHz that could correspond to the $J$ = 21-20 and $J$ =
25-24 transitions of CH$_3$C$_3$N respectively. The spectra
covering other 3 mm transitions of CH$_3$C$_3$N have not enough
sensitivity to confirm or discard the presence of this species.

The synthesis of CH$_3$CCH involves various ion-molecule reactions
with the dissociative recombination of the ions C$_3$H$_5^+$ and
C$_4$H$_5^+$ as the last step. The model underestimates the
observed CH$_3$CCH column density by one order of magnitude,
probably due to uncertainties and/or incompleteness in the
chemical network. The heavier chain CH$_3$C$_4$H is predicted to
have a column density even higher than that of CH$_3$CCH. The
smaller rotational constant B and larger rotational partition
function of CH$_3$C$_4$H makes it less favorable for being
observed at millimeter wavelengths. However its larger dipole
moment ($\mu_{CH_3C_4H}$ = 1.207 D versus $\mu_{CH_3CCH}$ = 0.784
D) could result in line intensities similar to those of CH$_3$CCH.
Our IRAM 30-m $\lambda$ 3 mm data shows an unidentified feature
with $T_A^*$ $\sim$3 mK at 81427 MHz that could be assigned to the
$J$ = 20-19 transition of CH$_3$C$_4$H. As occurs with
CH$_3$C$_3$N, the detection of CH$_3$C$_4$H remains tentative
since our current spectra covering other 3 mm lines is not
sensitive enough.

Lastly, as previously discussed by \citet{agu06}, H$_2$CS is
formed by the neutral-neutral reaction S + CH$_3$ and the
dissociative recombination of H$_3$CS$^+$. The predicted column
density is N$_{tot}$ = 1.3 $\times$ 10$^{12}$ cm$^{-2}$, which is
a factor 7 lower than the value derived from observations. We note
that a significant fraction of both H$_2$CO and H$_2$CS could be
formed on grain surfaces by hydrogenation of CO and CS
respectively. It is remarkable that in IRC +10216 thioformaldehyde
is more abundant than formaldehyde, despite the cosmic abundance
of oxygen being 50 times larger than that of sulphur. The
carbon-rich character of the CSE makes oxygen-bearing species to
have a very low abundance.

\section{Excitation mechanism: infrared pumping}

The chemical model predicts that the molecules formed in the outer
envelope have their peak abundances between 3 $\times$ 10$^{16}$
cm and 6 $\times$ 10$^{16}$ cm, which corresponds to angular radii
of 13''-26'' for an assumed distance to IRC +10216 of 150 pc (see
Fig.~\ref{fig-chem}). Interferometric observations at millimeter
wavelengths have located the molecular shell at about 15''
\citep{gue93,aud94,luc95} although the exact value depends on the
molecule and transition mapped. At this distance the gas density
is a few 10$^4$ cm$^{-3}$ and the gas kinetic temperature is lower
than 20 K \citep{ski99}. We may ask ourselves how the rotational
levels are excited to result in rotational temperatures for
CH$_2$CHCN and CH$_2$CN of 40-50 K, well above the gas kinetic
temperature.

Collisions with H$_2$ molecules do not seem to be responsible of
such an excitation. For example, in the case of the cyanomethyl
radical the collision coefficients $\gamma_{ul}$ for
CH$_2$CN-H$_2$ have not been measured or calculated but should be
similar to those calculated for CH$_3$CN-H$_2$ by \citet{gre86},
at least for the $K_a$ = 0 ladder (see \citet{tur90} for more
details). Thus, adopting a deexcitation collision coefficient
$\gamma_{ul}$ = 10$^{-10}$ cm$^3$ s$^{-1}$, typical for
$\Delta$$J$ = -1, $\Delta$$K$ = 0 transitions of CH$_3$CN, and an
Einstein coefficient for spontaneous emission of $A_{ul}$ = 5
$\times$ 10$^{-5}$ s$^{-1}$, the resulting critical density is
$n_{crit}$ = $A_{ul}$/$\gamma_{ul}$ = 5 $\times$ 10$^5$ cm$^{-3}$.
This value is larger than the gas density expected at a distance
of 4-5 $\times$ 10$^{16}$ cm, so that the rotational levels are
not thermalized and collisional excitation by itself should result
in a rotational temperature lower than the gas kinetic
temperature. In the cases of molecules such as CH$_2$CN, with a
relatively large dipole moment, i.e. whose rotational levels are
hardly thermalized in the outer envelope, and which have
rotational temperatures above the kinetic temperature of the gas,
the excitation is most probably dominated by a radiative
mechanism.

Absorption of infrared photons and pumping into excited
vibrational levels followed by radiative decay to rotational
levels in the ground vibrational state has been invoked many years
ago as the main excitation mechanism for some circumstellar
molecules in IRC +10216 \citep{mor75}. As a matter of fact,
several molecules present in the outer envelope have been detected
in excited vibrational states, e.g. C$_4$H $\nu_7$ = 1, 2 ($\nu_7$
= 131 cm$^{-1}$; \citealt{gue87,yam87}), HC$_3$N $\nu_7$ = 1
($\nu_7$ = 223 cm$^{-1}$; \citealt{cer00}), l-C$_3$H $\nu_4$ = 1
($\nu_4$ = 28 cm$^{-1}$; \citealt{cer00}) and SiC$_2$ $\nu_3$ = 1
($\nu_3$ = 160 cm$^{-1}$; \citealt{gen97,cer00}). Radiative
transfer calculations have also evidenced the importance of
infrared pumping in exciting molecules such as HC$_5$N
\citep{deg84} and H$_2$O \citep{agu06} in IRC +10216. For this
process to be efficient, the molecules must have vibrational modes
active in the infrared, sufficiently strong and with a frequency
at which the central object emits a high flux. The spectrum of IRC
+10216 as seen by ISO peaks at a wavelength of $\sim$10 $\mu$m
\citep{cer99} and the flux is very large within a wide wavelength
range (see below). In fact, the excited vibrational states of
C$_4$H, HC$_3$N, l-C$_3$H and SiC$_2$ detected have wavelengths
between 45 and 357 $\mu$m.

In the case of CH$_2$CHCN, there exists a large number of
vibrational modes, the strongest of which are the bending modes
$\nu_{12}$ and $\nu_{13}$ at 972 cm$^{-1}$ (10.3 $\mu$m) and 954
cm$^{-1}$ (10.5 $\mu$m) respectively (\citealt{cerceau85,khl99}).
For CH$_2$CN the only available experimental information concerns
the $\nu_5$ mode at 664 cm$^{-1}$ (15.1 $\mu$m; \citealt{sum96}).

In order to give a quantitative estimate of how important is the
infrared pumping compared to excitation by collisions, we may
evaluate the rates of both processes. The excitation by infrared
pumping operates through absorption of an infrared photon and
promotion of a molecule from a given rotational level in the
ground vibrational state ($v_0$, $J''$) to another rotational
level in an excited vibrational state ($v_1$, $J$), from which it
decays spontaneously to a different rotational level of the ground
vibrational state ($v_0$, $J'$) through either a single transition
or a radiative cascade process. The rate at which the latter level
($v_0$, $J'$) is populated is then governed by the rate of the
absorption process: $v_0$, $J''$ $\rightarrow$ $v_1$, $J$, which
is given by:
\begin{equation}
R_{\rightarrow v_0, J'} = B_{v_0 \rightarrow v_1} 4 \pi
\overline{J} \qquad [=] \quad s^{-1}
\end{equation}
where $B_{v_0 \rightarrow v_1}$ is the Einstein coefficient for
absorption and $\overline{J}$ is the mean intensity, averaged over
all solid angles, at the frequency of the infrared band. The
deexcitation from the level ($v_1$, $J$) can be assumed as
instantaneous in IRC +10216 for all relevant wavelengths.

The rate at which the rotational level ($v_0$, $J'$) is populated
by collisions with H$_2$ molecules is given by:
\begin{equation}
C_{\rightarrow v_0, J'} = \gamma_{lu} n(H_2) \qquad [=] \quad
s^{-1}
\end{equation}
where $\gamma_{lu}$ is the coefficient for rotational excitation,
within the ground vibrational state, by collisions with H$_2$ and
$n\rm (H_2)$ is the H$_2$ volume density.

\begin{figure}
\includegraphics[angle=0,scale=.47]{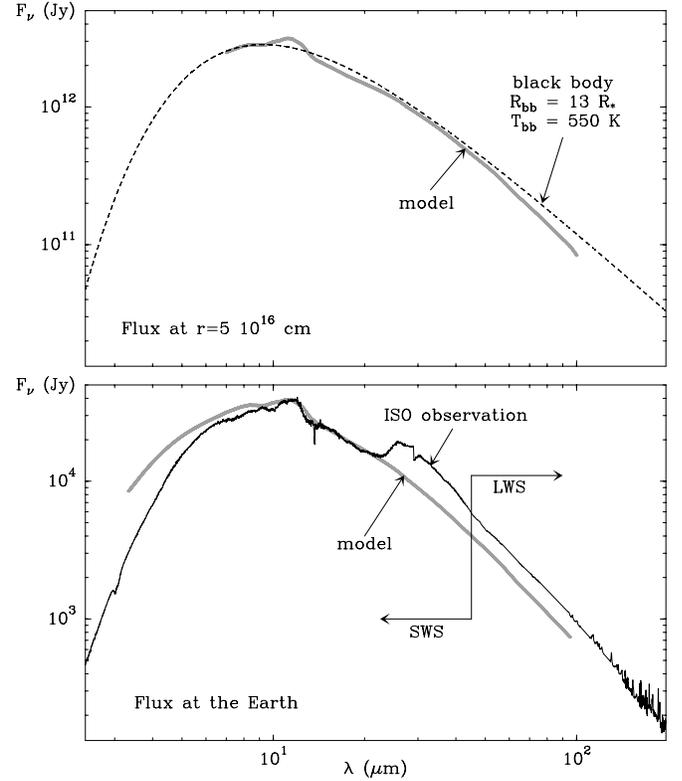}
\caption{Spectral energy distribution of IRC +10216 at infrared
wavelengths. The lower panel shows the fit to the continuum
observed by ISO. Most of the discrepancies between model and
observations in the 7-20~$\mu$m range are likely due to molecular
infrared bands, which are not considered in the model. In the top
panel the radiation field (4$\pi$$\overline{J}$) at a point 5
10$^{16}$ cm away from the star, as given by the model, is
compared to that expected with a central blackbody of radius 13
R$_*$ and temperature of 550 K.} \label{fig-cwleo-ir}
\end{figure}

To estimate the infrared flux in the outer envelope we have used
the radiative transfer model described in \citet{fon07} and the
ISO observations of IRC +10216 \citep{cer99}. The best fit to the
observed continuum between 7 and 27 $\mu$ is shown in the lower
panel of Fig.~\ref{fig-cwleo-ir}. The SiC band at 11.3 $\mu$m is
correctly reproduced. At long wavelengths the model seems to
underestimate the absolute observed flux, which is nevertheless
affected by the uncertainties in the multiplicative calibration
factors applied in the data reduction \citep{swi98}.

The model has been used to calculate the radiation field at a
distance of r = 5 $\times$ 10$^{16}$ cm (20$''$) from the star
(see upper panel in Fig.~\ref{fig-cwleo-ir}). The calculated
spectral energy distribution can be approximated by a blackbody
with a radius R$_{bb}$ = 13 R$_*$ and a temperature T$_{bb}$ = 550
K (see upper panel in Fig.~\ref{fig-cwleo-ir}), so that we may use
the Planck function for such a blackbody to evaluate
4$\pi$$\overline{J}$ and see whether the rate for infrared pumping
exceeds or not the collisional excitation rate. The radiative over
collisional ``excess'' ($R_{\rightarrow v_0, J'}$/$C_{\rightarrow
v_0, J'}$) may be evaluated as:
\begin{equation}
\frac{R_{\rightarrow v_0, J'}}{C_{\rightarrow v_0, J'}} =
\frac{(g_{v_1, J}/g_{v_0, J''}) A_{v_1 \rightarrow v_0}
(R_{bb}/r)^2}{4 [e^{(h \nu_{IR}/k T_{bb})} - 1] \gamma_{lu}
n(H_2)}
\end{equation}
where ($g_{v_1, J}$/$g_{v_0, J''}$) is the ratio between the
statistical weights of the ($v_1, J$) and ($v_0, J''$) states,
$\nu_{IR}$ and $A_{v_1 \rightarrow v_0}$ are the frequency and
Einstein coefficient of spontaneous emission of the vibrational
transition, and $h$ and $k$ are the Planck and Boltzmann
constants. Adopting, to a first approximation, ($g_{v_1,
J}$/$g_{v_0, J''}$) equal to unity and taking into account the
variation of the H$_2$ volume density with radius, $n\rm (H_2)$
[cm$^{-3}$] = 3.1 $\times$ 10$^{37}$ / $r^2$ with $r$ expressed in
cm (see e.g. \citealt{agu06}), we arrive at:
\begin{equation}
\frac{R_{\rightarrow v_0, J'}}{C_{\rightarrow v_0, J'}} =
\frac{5.8 \times 10^{-9} A_{v_1 \rightarrow v_0}(s^{-1})}
{[e^{(26.2/\lambda_{IR}(\mu m))} - 1] \gamma_{lu}(cm^3 s^{-1})}
\end{equation}
The radiative over collisional excess is independent of radius,
since both the radiation field and the H$_2$ volume density
decrease with the square of the radius, and only depends on the
wavelength and strength of the vibrational band and on the
collision coefficient. Just as an example, if we take typical
values $\gamma_{lu}$ = 10$^{-10}$ cm$^3$ s$^{-1}$ and $A_{v_1
\rightarrow v_0}$ = 1 s$^{-1}$ (adequate for vibrational modes
sufficiently strong), we obtain that the rate of infrared pumping
exceeds the collisional excitation rate for vibrational modes with
wavelengths $\lambda_{IR}$ larger than $\sim$ 6 $\mu$m. Bending
modes have usually wavelengths larger than this lower limit, thus
if the mode is active in the infrared and sufficiently strong, the
rotational levels of that molecule will be mostly excited through
infrared pumping in the circumstellar envelope of IRC +10216. For
example, the two bending modes of CH$_2$CHCN $\nu_{12}$ (972
cm$^{-1}$; 10.3 $\mu$m) and $\nu_{13}$ (954 cm$^{-1}$; 10.5
$\mu$m) have Einstein coefficients for spontaneous emission of 2.9
s$^{-1}$ and 2.8 s$^{-1}$ respectively\footnote{See e.g.
http://www.lesia.obspm.fr/~crovisier/basemole/}. The collision
coefficients $\gamma_{ul}$ for CH$_2$CHCN-H$_2$ are not known, but
according to equation (6), values as large as 5 $\times$ 10$^{-9}$
cm$^3$ s$^{-1}$ are required for making collisions to dominate
over infrared pumping. Thus, this molecule will be preferably
excited by infrared pumping through these two bending modes in IRC
+10216.

\section{Discussion}

\begin{figure*}
\includegraphics[angle=-90,scale=.74]{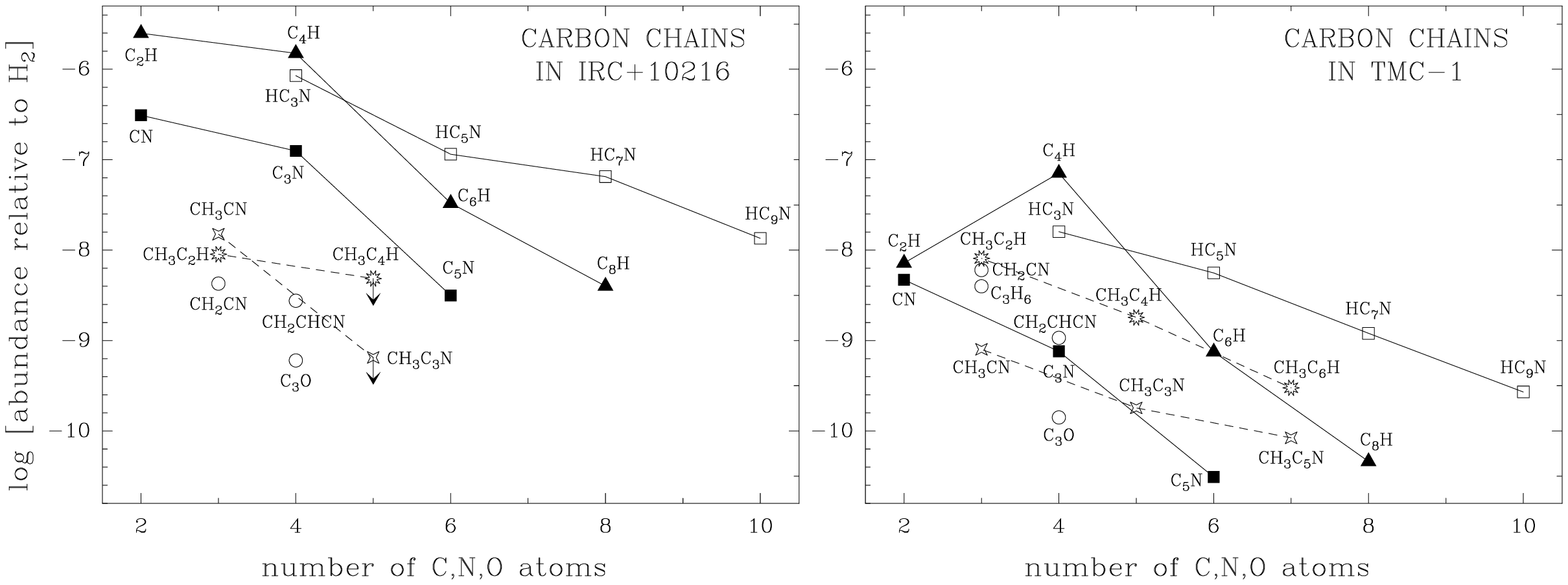}
\caption{Abundances of carbon chain molecules in IRC +10216 and
TMC-1. The diagram is an extension of that published by
\citet{cer87}. The fractional abundances relative to H$_2$ are
computed from the molecular column densities and the total H$_2$
column density. We use N(H$_2$) = 10$^{22}$ cm$^{-2}$ in TMC-1
\citep{cergue87} and N(H$_2$) = 2 $\times$ 10$^{21}$ cm$^{-2}$ in
IRC +10216. The latter value corresponds to the total H$_2$ column
density contained in an outer shell extending from 2 $\times$
10$^{16}$ cm to 7 $\times$ 10$^{16}$ cm, where all the molecules
considered in the diagram are most probably present. The choice of
these boundaries is not critical for the values of the molecular
abundances. Only if any of the molecules considered does not form
in the outer envelope but in the inner regions of the CSE, then
its abundance would be substantially overestimated.}
\label{fig-irc-tmc1}
\end{figure*}

The detection in IRC +10216 of the partially saturated organic
molecules reported in this Paper invites for a comparison to
TMC-1, the other Galactic source that displays, together with IRC
+10216, the largest wealth in unsaturated carbon chain molecules.
For that purpose we have represented in Fig.~\ref{fig-irc-tmc1}
the abundances of various carbon chains in these two sources as a
function of the number of heavy atoms.

Chemistry proceeds in a different way in these two objects. In IRC
+10216 interstellar UV photons drive the chemistry by
photodissociating the closed shell molecules flowing out from the
inner envelope. In TMC-1 the gas is well shielded against the
interstellar UV field and the built-up of molecules is driven by
ionization by cosmic rays. This introduces a sharp difference in
the time scales of the processes being at work. In IRC +10216 the
gas travels throughout the CSE during a few thousands of years and
the chemical processes take place in shorter time scales. In TMC-1
the built-up of molecules has time scales of the order of millions
of years and the depletion from the gas of some species, due to
condensation on grains, is a major issue at late times. A
consequence of this is the lower level at which molecules are
present in TMC-1 compared to IRC +10216. As an example we see in
Fig.\ref{fig-irc-tmc1} that the most abundant carbon chains in
both sources are those with a highly unsaturated character,
cyanopolyynes and polyyne radicals, the abundances of which are
larger in IRC +10216 than in TMC-1 by more than one order of
magnitude.

Another important difference between these two sources could lie
in the C/O ratio. In IRC +10216, the processes of dredge-up
related to the AGB phase have resulted in a photospheric C/O ratio
higher than 1. This controls the chemistry in the outer envelope:
almost all the oxygen keeps locked into CO during most of the
expansion while the carbon in excess can participate in a rich
carbon chemistry. In TMC-1 it is not clear whether the C/O ratio
is higher or lower than 1. The large number of carbon chains found
argues in favor of a C/O ratio $>$ 1, which is unusual for a dense
cloud, and some chemical models have used it because a better
agreement with observations is achieved \citep{smi04}. On the
other hand, oxygen is cosmically more abundant than carbon and
TMC-1 contains a substantial number of O-bearing molecules
\citep{ohi98}, which points toward a C/O ratio $<$ 1. If the
latter is true, then the carbon chemistry would be limited due to
the presence of atomic oxygen which tends to destroy the chemical
complexity \citep{her94}, something that does not happen in IRC
+10216 due to the unavailability of free atomic oxygen
\citep{mil94}. This could also explain the systematically lower
abundances of carbon chains in TMC-1 compared to IRC +10216.

If we focus on partially saturated species, such as
methylpolyynes, methylcyanopolyynes, CH$_2$CN or CH$_2$CHCN, then
we note that in IRC +10216 they are present at a lower level than
the highly unsaturated species, while in TMC-1 partially saturated
and highly unsaturated molecules have abundances in many cases
comparable; e.g. compare the ratio of the abundances of CH$_2$CN
and HC$_3$N in both sources. Thus, chemistry seems to favor the
presence of species with a higher degree of saturation in dark
clouds than in C-rich circumstellar envelopes. The recent
detection of the highly saturated hydrocarbon propylene in TMC-1
\citep{mar07} supports this point.

In IRC +10216 the UV field which drives the chemistry is mostly
photodissociating but not ionizing, so that the dominant reactions
are neutral-neutral rather than ion-molecule. In TMC-1, the
chemistry is triggered by cosmic rays ionization, and ion-molecule
reactions do greatly participate in the built-up of molecules.
Note that chemical models have traditionally explained the
formation of molecules in interstellar clouds by means of
ion-molecule reactions. However, since it was discovered that some
neutral-neutral reactions are very rapid at low temperatures, it
is now thought that these latter reactions dominate the formation
of highly unsaturated carbon chains such as cyanopolyynes
\citep{her94}. That is, in general terms we may state that
``current gas phase chemical models form highly unsaturated carbon
chains mostly by neutral-neutral reactions while partially
saturated molecules are generally formed by ion-molecule
reactions''. Therefore, ion-molecule reactions could be,
ultimately, the responsible of the higher degree of saturation
observed in TMC-1 compared to IRC +10216. Grain surface reactions,
which are very efficient in producing highly saturated molecules,
could also play a role in the peculiar chemistry of TMC-1.
However, it is not clear how these mantle species would be
desorbed to the gas phase at the low temperatures prevailing in
these regions.

\section{Conclusions}

In summary, we have shown that apart from the abundant highly
unsaturated cyanopolyynes and polyyne radicals, analogous
molecules with a higher degree of saturation are also present in
IRC +10216. In this paper we have presented the detection of
CH$_2$CHCN, CH$_2$CN, CH$_3$CCH and also of thioformaldehyde. The
relatively high rotational temperatures, 40-50 K, derived for
CH$_2$CHCN and CH$_2$CN suggest that these species are excited in
the circumstellar envelope through radiative pumping to excited
vibrational states.

The formation of partially saturated organic molecules in IRC
+10216 resembles that occurring in cold dense clouds and stresses
the similarity between the chemistry in these two types of
sources. However, unlike in TMC-1, their abundances are much lower
than those of highly unsaturated molecules, like cyanopolyynes and
polyyne radicals, which reflects the differences in the chemical
processes at work in dark clouds and C-rich circumstellar
envelopes.

\begin{acknowledgements}

This work has been supported by Spanish MEC trough grants
AYA2003-2785, ESP2004-665 and AYA2006-14876, by ``Comunidad de
Madrid'' under PRICIT project S-0505/ESP-0237 (ASTROCAM) and by
the European Community's human potential Programme under contract
MCRTN-CT-2004-51230 (The Molecular Universe). MA also acknowledges
funding support from Spanish MEC through grant AP2003-4619.

\end{acknowledgements}

\end{document}